\documentclass[pre,aps,superscriptaddress,fleqn,showpacs,twocolumn]{revtex4}
\usepackage{amsmath,amssymb,multirow,graphicx}
\begin{document}

\title{Druse-Induced Morphology Evolution in Retinal Pigment Epithelium }

\author{K.~I.~Mazzitello}

\affiliation{CONICET,
             Universidad Nacional de Mar del Plata, 7600 Mar del Plata,
         Argentina}

\author{Q.~Zhang}
\affiliation{Department of Ophthalmology, 
	Emory University, Atlanta, GA 30322, USA}

\author{M.~A.~Chrenek}
\affiliation{Department of Ophthalmology, 
	Emory University, Atlanta, GA 30322, USA}

\author{F.~Family}
\affiliation{Department of Physics, Emory University, Atlanta, GA 30322,
             USA}

\author{H.~E.~Grossniklaus}
\affiliation{Department of Ophthalmology, 
	Emory University, Atlanta, GA 30322, USA}

\author{J.~M.~Nickerson}
\affiliation{Department of Ophthalmology, 
	Emory University, Atlanta, GA 30322, USA}

\author{Y.~Jiang}
\affiliation{Department of Mathematics and Statistics, 
	Georgia State University, Atlanta, GA 30303,USA}

%\date{\today}  

\begin{abstract}

The retinal pigment epithelium (RPE)  is a key site of pathogenesis for many retina diseases.  The formation of drusen in the retina is characteristic of retinal degeneration. We investigate morphological changes  in the RPE in the presence of soft drusen using an integrated experimental and modeling approach.  We collect RPE flat mount images from donated human eyes and develop 1) statistical tools to quantify the images and 2) a cell-based model to simulate the morphology evolution.  We compare three different mechanisms of RPE repair 
evolution, cell apoptosis, cell fusion, and expansion, and 
Simulations of our RPE morphogenesis model quantitatively reproduce deformations of human RPE morphology due to drusen, suggesting that a purse-string mechanism is sufficient to explain how RPE heals cell loss caused by drusen-damage. We found that drusen beneath tissue promote cell death in a number that far exceeds the cell numbers covering the drusen. Tissue deformations are studied using area distributions, Voronoi domains and a texture tensor.

\end{abstract}
 \maketitle

\section{Introduction}
 
Age-related macular degeneration (AMD) is a degenerative disease of the photoreceptors as well as the retinal pigment epithelium (RPE), which form a key layer in the central part of the human retina called the macula \cite{ardeljan2013}. Early stages of the disease feature deposition of extracellular debris, known as drusen, from the basal side of the RPE onto Bruch's membrane.  The later stages of the disease may either progress to one of two forms, known as geographic atrophy (dry AMD) and choroidal neovascularization (wet AMD). Patients with dry AMD exhibit loss of photoreceptors and RPE in the macula, hence loss of the central vision. Patients with choroidal neovascularization experience abnormal growth of choroidal vasculature that cross Bruch's membrane and in some cases RPE and the neuroretinal layers.  These poorly regulated, abnormal vessels leak serum and blood into the retina and can cause blindness  \cite{Coleman2008, cite3Rabbit07}. Late-stage of AMD is the  leading cause of  blindness in adults beyond the age of 55 \cite{Congdon2004, Pascolini2004, citetoMiller20, cite1Miller10, Pascolini2011} and is projected to afflict more than 3 million Americans by the year 2020 \cite{Friedman2004}.  Clinically there has been moderate success with intravitreal injection of anti-vascular endothelial growth factor (VEGF) to deter choroidal neovascularlization \cite{Brown09, Forte10}, but there are no robust methods to treat patients suffering from geographic atrophy.  The long-term prognosis of this form of AMD is poor in many cases, mainly because there is no way to differentiate normal aging eye from the eye with early stages of dry AMD. Any progress in differentiating between a normal eye and one with early stages of AMD will have a major impact on the development  new therapies as well as new tools for diagnosing different stages of AMD. \par  

\begin{figure}[!ht]
\begin{center}
\includegraphics[width= 8 cm]{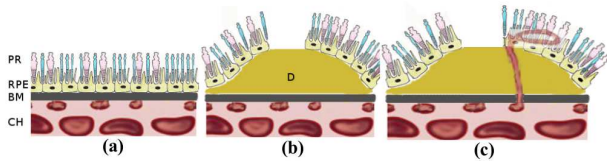} 
\end{center}
\caption{Illustrations of the retinal pigment epithelium (RPE) of the retinal pigment epithelium (RPE) layer,  Bruch's membrane (BM), and the choroid region histopathology associated with various macular degenerative processes (adapted from \cite{Song2013} with permission): (a) normal macula, (b) with a druse sandwiched between the RPE and BM with adjacent geographic RPE atrophy and loss of overlying photoreceptors (PR), and (c) choroidal neovascularization is the abnormal growth of new blood vessels in the choroid (CH) that penetrate  the BM and the RPE. }
\label{fig1}
\end{figure}

Figure \ref{fig1}(a) is a schematic illustration of the region of the macula involved in AMD. Choroidal neovascularization (CNV) is the abnormal growth of blood vessels through Bruch's membrane and the RPE layer from the choroid region. CNV can occur in the sub-RPE, the sub-retinal space, or both (Fig. \ref{fig1}(c)). These abnormal vessels leak serum and blood that can induce a fibrotic reaction known as a disciform scar.  The modern-day therapies rely mainly on anti-angiogenic drugs that they have been a great success in treating CNV \cite{Brown2008, Martin2011}. The RPE layer, situated between the choroidal vasculature region and the photoreceptors, is arguably key site of AMD pathology. RPE is composed of a single layer of cells (Fig. \ref{fig2}). This layer has several functions including participation in the regeneration of 11-cis-retinal in the visual cycle, absorption of spray light passing through the retina, formation of the outer blood-ocular barrier, upkeep of subretinal space including fluid and electrolyte balance, maintenance of the choriocapillaris, and the phagocytosis of shed photoreceptor outer segments. Healthy RPE cells are critical for maintaining the structure of the retina and preserving normal photoreceptor function. In addition to AMD, abnormal RPE cells contribute to the formation and progression of numerous retinal diseases, including Stargardt's dystrophy and Best's disease \cite{williams09}. 

%%%%%%%%%%%%%%%%%%%%%%%%%%%Figure 1 %%%%%%%%%%%%%%%%%%%%%%%%%%%%%%%%%%%%%

We have previously showed that RPE morphology can discriminate the age and disease status of the eye in mouse models \cite{jiang2013}. Our goal in the present paper is to determine how drusen modifies RPE morphology as AMD progresses. \par  

As the eyes age, intracellular and extracellular debris start to accumulate in the retina and mediate macular dystrophies. For example, excess accumulation of autofluorescent lipofuscin pigments, that are composed of finely granular yellow-brown pigment granules that are made  of lipid-containing residues of lysosomal digestion, in the RPE have been shown to be cytotoxic in multiple studies and a major contributor to degenerative diseases including macular degeneration \cite{Sparrow2005, Winkler1999}.  Drusen are small yellow or white accumulations of extracellular material that build up between Bruch's membrane and the RPE.   The origin of the proteins and lipids in drusen is not entirely known, but drusen can be considered as ocular landfills \cite{ardeljan2013}.  Drusen typically fall under three types. Hard drusen appear to be yellow dots, usually smaller than 30 $\mu m$ in radius, and have sharp borders \cite{Millodot}. Soft drusen, which are not considered a part of normal aging, are larger, typically greater than 62 $\mu m$ in radius, and may have either sharply defined borders or fuzzy, are mostly aggregates of hard drusen that are formed over time \cite{Millodot}. Basal linear deposits are composed of primarily membranous material in the inner collagenous layer of Bruch's membrane and are fibrous and amorphous and located between RPE and its basal lamina \cite{Green1999}. Of these different types of deposits, only hard drusen are considered to be a consequence of normal aging. They induce a lifting the RPE layer, stretching and sometimes compressiong cells without evidence of cell death. The presence of hard drusen does not significantly increase the risk of developing AMD \cite{Buch2005}, but other deposits, particularly soft drusen, have been implicated in AMD \cite{Rudolf2008}. The early presence of soft drusen is considered a great risk factor of CNV and early loss of vision. Here we focused on the study of deformations of the RPE morphology triggered by soft drusen. \par 

Soft drusen cause morphology changes in RPE cells. Like other transporting epithelia, RPE cells contain distinct apical and basal-lateral plasma membrane domains separated by adherent complexes that encircle the cell at the boundary between apical and lateral regions \cite{Mostov02, Nilsson, Mawas, Moyer}. Much of the molecular machinery that reorganizes epithelial contacts is known.  The elements of the complex, identified as zonula occludens (tight junctions), zonula adherens (intermediary junctions), and macula adherens (desmosomes), occupy a juxtaluminal position and succeed each other in the order given in an apical-basal direction \cite{Farquhar}. The adherens junctions and desmosomes, together with the tight junctions form junctional complexes that completely encircle the cells, resulting in a continuous junctional belt that interconnects neighboring cells \cite{Farquhar}, offering both adhesive and contractive forces.  Modulation of the adhesion and contraction of cell contacts changes the morphology of the junctional network, and contributes to morphogenetic movements in epithelial sheets \cite{Julicher2007:3, Julicher2007:10, Julicher2007:11}. To understand RPE morphogenesis and changes resulting from druse-induced cell death, we need a physical description of the epithelium that can account for contractile and adhesive forces of the cells and tissue remodeling of the cells in tight connections. For this purpose, we developed a 2D epithelium tissue model that accounts for the damage and recovery of cells through a purse-string mechanism in which wounds closed as if the entire wound border were subject to circumferential tension and besides investigated the epithelial morphology changes as a function of druse size. Although multi-nucleate RPE cells have been observed that may arise due to endoreplication, cell fusion, or incomplete cell division \cite{Starnes2016}, we proposed here that apoptosis followed by purse-string epithelial repair is the preponderant mechanism.  However, to validate our hypothesis, we used computer simulations comparing cell patterns damaged by drusen beneath the tissue in which cells die by apoptosis, cell fusions or expansions followed by the death of neighbor cells due to an excessive compression.\par

%It was observed that wounds closed smoothly in a purse string fashion, as if the entire wound border were subject to circumferential tension.
%Se observó que las heridas se cerraron sin problemas de una manera cordón de bolsa, como si toda la frontera herida estaban sujetos a tensión circunferencial.

\section{Materials and Methods}
We obtained and analyzed human RPE flat mount tissues. In particular we focused on drusen induced cell morphology. The study protocol adhered to the tenets of the Declaration of Helsinki for research involving human subjects, including identifiable human tissue. Donated eyes with AMD and age-matched controls were obtained from the Georgia Eye Bank, Atlanta, GA, within 48 hours of death and stored in moist chambers at 4 $^{\circ}$C after enucleation. The globes were fixed in 10\% buffered formalin (pH 7.4) for 4 hours and stored in 1x Phosphate Buffered Saline, pH 7.4 (PBS, Invitrogen Corporation, Carlsbad, CA) at 4 $^{\circ}$C. The RPE flat-mounts were prepared using a microdissection technique as follows. Briefly, the enucleated eyes were placed in a plastic ``egg-cup" support under a dissecting microscope. A scleral incision 3 mm posterior to the limbus was made using a disposable blade, extended 360 degrees circumferentially parallel to the limbus. The cornea, iris and lens was removed. Six radial anterior-posterior oriented scleral cuts were placed at the the ora serrata and extended towards the optic nerve to enable the tissue to be flattened. The star-shaped flatmounts were left during scleral limbal trimming and placed on a glass slide. The scleral petals were carefully peeled away to expose the intraocular tissue with particular attention paid to the vortex veins and fovea. The retina and vitreous were subsequently removed in one piece by gentle teasing away of the choroid-RPE sheet from the ciliary insertion with the specimen submerged in PBS buffer. The flatmount preparation was divided into six portions (temporal, inferotemporal, inferonasal, nasal, superonasal, and superotemporal). After dissection, the tissues were flat-mounted, RPE uppermost, onto paper-frame bordered glass slides. The specimens were rinsed with PBS followed by staining with 2.5\% AF635-phalloidin (Invitrogen, Camarillo, CA) and 20 ug/ml propidium iodine in 0.1\% Triton X-100 HBSS solution. Subsequently, the specimens were washed 3 times with 2x 0.1\% Triton X-100 HBSS buffer, mounted with 2 drops vectashield hardset, coverslipped, and allowed to set overnight. Images were renditions of 3 optical sections each spaced $5 \mu m$ apart in Z-stacks and were imaged with a confocal microscope. Each confocal image was 1024x1024 pixels in size and the green channel was used for our analysis. In the confocal images the green, red and blue corresponded to AF635-phalloidin staining of actin cytoskeleton, propidium iodide staining of nuclei and autofluorescence, respectively.

\section{Cellular Potts Model of RPE morphogenesis} 
\label{model}

Many mathematical models have focused on the study of embryonic epithelial morphogenesis (e.g. \cite{Odell}), wound healing (e.g. \cite{Murray, McDougall, Posta}) and more recently on disruption of epithelial homeostasis and its connection to cancer (e.g. \cite{Shin}). 

%Other mechanisms in the RPE cells were observed, such as fusion cell, but the purse-string mechanism preponderates.
% RPE PATTERN A snapshot of dynamics -- want to decipher the underlying mechanisms leading to the pattern
% in Drosophila wing (elife paper) 

A challenge in mathematical modeling is the level of detail necessary for the description of a biological system under study, particularly whether single cell behavior needs to be included in the description of the system or not.   Although for some systems the gene networks regulating embryonic patterning have been identified in detail \cite{Merks09:12}, descriptions of gene networks are rarely directly involved in the dynamics of cell behavior, including cell motility, cell adhesion, and chemotaxis. Therefore mathematical models are typically based on experimentally plausible, qualitative descriptions of cell behavior. We used the Cellular Potts model \cite{Merks09:16}, which has been used in many studies of biological phenomena and is an effective cell-based approach for modeling biological growth and morphological development (for a  review of the Cellular Potts model approach, see \cite{Merks09:33}; for more recent applications, see \cite{Merks09:21, Merks09:18, Merks09:38, Merks09:33, Merks09:58}  ). \par

In \cite{Graner2015} a multiscale novel formalism was proposed, in which relates the characterizations of each cell process (cell divisions, cell rearrangements, cell shape and size changes, apoptoses, cell fusions, etc) to tissue growth and morphogenesis. Based on a texture tensor, the formalism unambiguously measures the tissue deformation rate as well as the deformation rates associated with each individual cell process. The authors have used the Cellular Potts model to validate the formalism and illustrate the impact of cell divisions on tissue elongation and on the other processes. There is a complex interplay between cell divisions and other processes such as cell shape changes and rearrangements. They found that the cell divisions are not always responsible of the elonngations that suffer a tissue during its mophogenesis as it is exposed in the literature. The formalism has been theoretically demostrated \cite{Graner2015}, then in a sense validates the Cellular Pott model as an excellent model to study cell rearrangements and cell size and shape changes on tissues.

We used a two-dimensional (2D) Cellular Potts model for the RPE morphogenesis.  This choice is appropriate because the RPE is essentially a 2D sheet. Moreover, the adherent complexes that we stained to identify each cell in experiments, be it the tight junction or actin, were distributed on the cell's edges on the apical face. The Cellular Potts model is now a standard tool to simulate multiple cell patterning based on single cell behaviors \cite{25ofGraner07}. This model represents most cell behaviors in the form of a generalized energy $E$, which includes the interactions between cells, between a cell and its cellular environment and constraints that determine individual cell behaviors.  One of the motivations for its application in biology is its capability to handle irregular, fluctuating interfaces. Examples that illustrate its capabilities are modeling of cell sorting in aggregates of embryonic chicken cells \cite{Glazier95}, morphological development of the slime mold {\it Dictyostelium} \cite{Jiang98}, avascular tumor growth \cite{Jiang05}, and choroidal neovascularization in the retina of AMD eyes \cite{Jiang2012}.  In the Cellular Potts model, a cell consists of a fluctuating domain of lattice sites, thus describing cell volume and shape more realistically. This spatial realism is important when modeling interactions dependent on cell geometry. Thus, the pixelization induced by the calculation lattice can be chosen to correspond to the pixelization in experimental images. This approximation allows the use of experimental images as initial configurations for reproducing morphological changes of cell rearrangement. We take advantage of this characteristic of the Cellular Potts model to study morphological changes of RPE associated with cell loss by druse formation.

%%%%%%%%%%%%%%%%%%%%%%%%% figure 2 %%%%%%%%%%%%%%%%%%\includegraphics[width=.8\textwidth]
\begin{figure}%[!ht]
\begin{center}
\includegraphics[width=8 cm ]{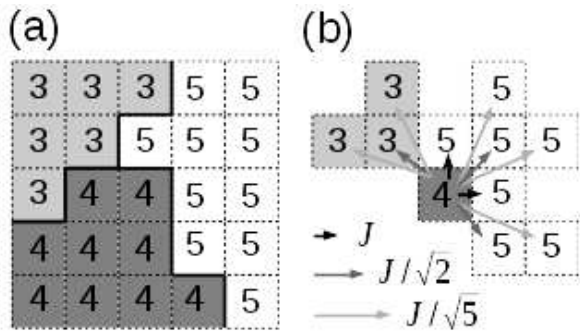} %&
\end{center}
\caption{Schematic illustration of the Cellular Potts Model: (a) Extract of a few sites of a square lattice.  Each lattice site has assigned a number and the cells are defined as domains of the same number. The solid line represents the cellular membrane. (b) The central site of the (a) extract and its nearest neighbor interactions.  The different weights for the different order of neighbors are indicated by arrows (darker arrow has stronger interaction), thus tending to form hexagonal cells (there is no interaction for the third order). Note that the sites labeled with the number $4$, except the central one, were not considered in (b) due to there is no interaction between sites belonging to the same cell (see first term of Eq.(\ref{energy})).}
\label{mod}
\end{figure}
%fourth nearest neighbor interaction, with different weights for the different
%order of neighbors

The details of Cellular Potts model have been presented on many occasions.  Briefly, a 2D Cellular Potts model describes packing geometries of epithelial cells as a 2D network.  Cells are defined as domains of the same numbers on a lattice. For convenience, we use a square lattice, but it is possible to use other kind of lattices according to the problem to be solved.  Each lattice site $i$ has assigned an integer cell number $\sigma_i$ chosen from ${1, ..., Q}$, where $Q$ is the total number of cells. The simple connected set of all sites with the same number $\sigma_i$ defines a cell, while links between different cell numbers define the cellular membrane (see \ref{mod}(a)). Thus, each cell number $\sigma_i$ simply acts as a label for a given cell extended over many lattice sites and their number defines the cell area $A_{\sigma_i}$. The total energy of a configuration is given by

\begin{eqnarray}\label{energy}
E=\sum_{ i,j\;neighbors} J_{ij} \left( 1-\delta_{\sigma_i\sigma_j} \right) +
\sum_{\sigma_i=1}^Q \kappa \left( A_{\sigma_i} - A_{\sigma 0}\right)^2
\end{eqnarray}
%where the summations are taken over occupied NN sites.
In this equation, $J_{ij}$ is the coupling strength between neighboring cells $\sigma_i$ and $\sigma_j$, summed over the entire lattice. $\delta_{\sigma_i\sigma_j}$ is the Kronecker delta, when $\sigma_i$ and $\sigma_j$ are not equal, the neighboring sites $i$ and $j$ belong to neighboring cells, effectively describing the interface between two cells. The first term of the energy (\ref{energy}) is associated with two cellular properties:  the cell-cell adhesion and the cortical tension of the cells. Then, adhesion and cortical tension have opposite effects on intercellular surface tension and on the extent of cell contact. The cellular adhesion is due to cadherin and junctional complex binding between cells. The cortical tension is due to the formation of contractile actin-myosin network at the zone of contact between cells \cite{Lecuit}. For two adhering contacting cells, the increase of this tension reduces the contact surface. They are not independent because both are supported by actin filaments \cite{Lecuit} and therefore their effects are considered together in the first term of Eq. (\ref{energy}).  The second term describes the elastic-area constraint with elastic constant $\kappa$, 
where $A_{\sigma_i}$ is the area of cell labeled with $\sigma_i$ and $A_{\sigma 0}$ is its preferred area, fixed in a range of sizes chosen using known biological data (I. E., $A_{\sigma 0}$ has a value associated with a normal or atypical cell area, subject to the situation studied, as will be seen at the end of this section).\par

We consider a fourth nearest neighbor interaction, with different weights for the different order of neighbors: for the first order neighbors, (0,1) and others according to lattice symmetry: $J_{ij}=J$; $J_{ij}=J/\sqrt{2}$ for the second order ((1,1) and symmetry); for the third order, $J_{ij}=0$; and for the fourth order, $J_{ij}=J/\sqrt{5}$, where $J$ is constant (see Fig.\ref{mod}(b)). The chosen interaction form effectively eliminates the energy anisotropy of the square lattice.\par  

The system evolves using Monte Carlo dynamics. Our algorithm differs from the standard Metropolis algorithm: we first randomly choose a lattice site that is at the cell boundary; we then propose that the site assumes one of its neighbor's ID value (I. E. a $\sigma_i$ value of a neighbor cell), and accept this change with probability $P$ given by the Boltzmann distribution, namely

\begin{eqnarray}\label{prob}
P=\left\{
\begin{array}{ll}
exp\left(-\Delta E/T\right) & \mbox{if}\;\;\Delta E>0  , \\
1   			    & \mbox{if}\;\;\Delta E\leq 0  .\\
\end{array}
\right.
\end{eqnarray}
Thermal shape fluctuations of the membranes on nm scales in
general
\noindent For the probability $P$, $\Delta E=\left(E_{final}-E_{initial}\right)$ is the energy change due to the proposed cell number update, and $T$ simulates cell membrane fluctuations governed by the cytoskeletal elasticity.
Higher $T$ implies higher fluctuations of the membrane.
Thermal fluctuations are much smaller than membrane fluctuations driven by cytoskeleton and therefore are not considered \cite{Glazier93}. The former ocurr on nm scales in general while the last ones ocurr on lenghscales greater than 100 nm \cite{Weikl2009}. Returning to the Eq. (\ref{prob}), a Monte Carlo step consists of as many cell number updates as there are lattice sites.  We use the Monte Carlo step (MCS) as the unit of time in what follows, but this definition of time unit is not directly related to real time \cite{16deGraner05}. If we let the system evolve to reach the lowest-energy configurations, a stationary and stable cell packing with cell membrane fluctuations is obtained. In the next section, we used large enough final times to obtain stable cell packing that simulateed normal cell patterns. These patterns satisfied a mechanical force balance where the total force vanishes. However, local changes in extracellular tissue, such as drusen formation or cell apoptosis, can produce alterations of cell forms \cite{Ingber:92}, and therefore configurations with higher energies characterized by local tensions can be found.\par

We want to distinguish between several possible mechanisms that may occur during the changes generated
on the RPE due to drusen. They are apoptosis followed by purse-string epithelial repair, cell fusions
which lead to bi- tri- up to five nucleate cells and cell expansions that can reach cell areas
five times larger that their original areas followed by the death of neighbor cells due to an
excessive compression.

Odell et al. \cite{Odell} first modeled epithelial tissue folding during development using a ``purse-string" idea: a subcortical band of microfilaments from each cell generates a force by shortening and driving the cell to change its shape, much like the drawing of a purse-string. The classic example of the purse-string is the well-studied process of dorsal closure in Drosophila \cite{Jacinto}, where the assembly and contraction of a multicellular acto-myosin belt lining the gap, aka purse-string \cite{Bement} is controlled by RhoA and its direct regulators Rho kinase (ROCK) and myosin light chain kinase (MLCK) \cite{Nobes}. We adopt this purse string model to simulate the closure of RPE sheet after cell death occurs by apoptosis because of soft drusen. 

\begin{figure}%[!ht]
\begin{center}
\includegraphics[width=8 cm ]{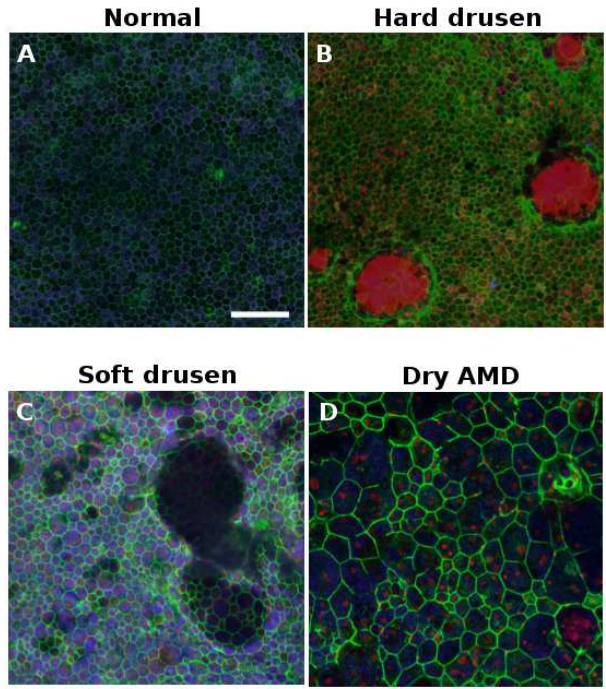} %&
\end{center}
\caption{Flatmount image of human RPE.  Actin was stained green with Alexa Fluor 635-Phalloidin, cell nuclei were stained red with Propidium iodide, Bule-violet indicates natural autofluorescence of slowly accumulating A2E-like derivates. Size bar indicates 100 $\mu m$. (A) Normal RPE morphology from an 88 year old female.  (B) Examples of hard drusen from the periphery of an 86 year old male with dry AMD.  Note the red color of the hard drusen indicating a high level of autofluorescence.  (C) Examples of soft drusen from near the macula of a 77 year old male with geographic atrophy.  Note the displacement of the RPE cells due to the drusen and lack of autofluorescence in the soft drusen. (D) Example of dry AMD progressing towards geographic atrophy in an 86 year old male with dry AMD.}
\label{fig2}
\end{figure}

To simulate the purse-string closure of a tissue damage after a localized cell death, we started with a digital normal RPE pattern (Fig. \ref{fig2}(a)) and converted its pixels to ID values on the lattice (Fig.\ref{mod}(a)). We assigned the measured current cell areas to be their target areas.  Cell apoptosis was modeled by setting the target volume of the cell to zero: $A_{dead\;cell\;0}=0$. At the same time, all the cells contacting this dead cell (nearest neighbors or NN), can enlarge according to $A_{1^{st}_{NN}\;0}=A_{1^{st}_{NN}\;0}+A_{dead\;cell}/$ (number of $NN$ cells), so the $NN$ cells stretched to over the area of the dead cell. The pattern evolves according to Eqs. (\ref{energy}) and (\ref{prob}) until the system reaches its lowest-energy configuration.\par 

The following process was used in the simulation of soft druse: Starting from a normal PRE pattern, we assigned a soft druse location and size; a cell is chosen at random from the soft druse area, and assigned to undergo apoptosis, and its nearest neighbors would then stretch to fill its space; this single cell apoptosis was repeated from within the soft druse area, until the pattern reached a similar cell morphology to that observed experimentally (see next section).\par

 \section{Results}

\subsection{Flat mount images of RPE with drusen}
 
We have obtained over 35 AMD eyes and 28 age-matched normal eyes from the eye bank.  Only those flat mounts with images that were good enough for analysis were used in this study.  Besides flat amount images of drusen with RPE, additional histological sectioning was performed on the same eyes to determine if the drusen were soft or hard.  

%%%%%%%%%%%%%%%%%%%%%%%%% figure 3 %%%%%%%%%%%%%%%%%%

\subsection{Morphological changes of RPE associated with soft drusen} 
\label{soft-drusen}
\begin{figure}%[!ht]
\begin{center}
\includegraphics[width=8 cm ]{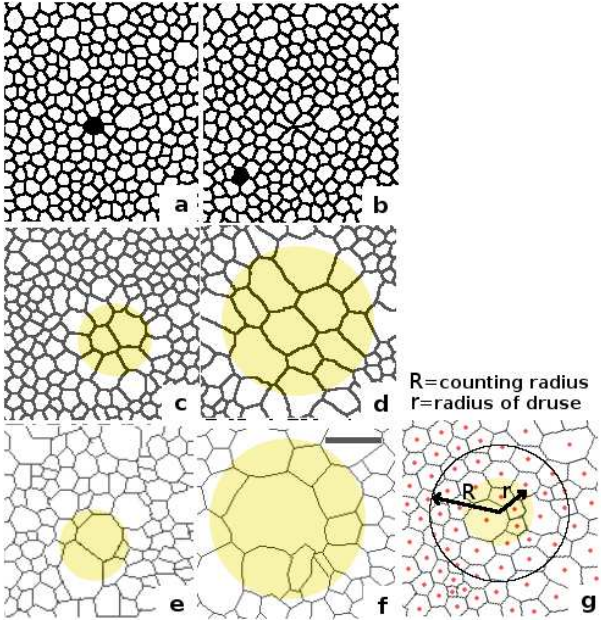} %&
\end{center}
\caption{RPE morphology in the presence of a soft druse: snapshots of simulations (a)-(d) and experiments (e)-(f). 
Shaded regions correspond to a soft druse.  (a) and (b) correspond to a simulated sequence of cell death and elongation of its neighbor cells. Black cells die and their first neighboring cells elongate to fill the area of dead cells. 
(c) and (d) are two final RPE images in the presence of soft druse of radius $35\;\mu m$  and $85\;\mu m$, respectively. 
(e) and (f) are traced, flatmount images showing only the cell boundary, from the macular region of the left eye of the same 87 year old male with geographic atrophy in Fig \ref{fig2}C. Scale bar = 50 $\mu m$. (g) is an illustration of the definitions of the radius of druse and radius of measurement.}
\label{snap}
\end{figure}

Empirical evidence suggested that the irregularities of tissue created by dying or dead RPE cells on a soft druse are repaired by the neighboring healthy RPE cells stretching and spreading to fill to gaps.  We modeled this repairing process (see section \ref{model}), allowing cellular rearrangements without cell mobility. These dynamics of deterioration take years in a human eye. Snapshots of simulations are shown in Fig. \ref{snap}. Black cells die and their first neighboring cells elongate to fill the area formerly occupied by the dead cell (snapshots (a) and (b)).  Comparisons among simulations (Fig. \ref{snap}c and Fig. \ref{snap}d) and RPE images with a soft druse of $35 \;\mu m$  of radius (Fig. \ref{snap}e) and $85 \;\mu m$ (Fig. \ref{snap}f)  show good qualitative resemblance. We further showed that the patterns were quantitatively similar. \par

%%%%%%%%%%%%%%%%%%%%%%%%%%%%%%%%%% FIGURE 4 %%%%%%%%%%%%%%%%%%%%%%%%%%%%%%%%%%%%%%

%%%%%%%%%%%%%%%%%%%%%%%%%%%%%%%%%% FIGURE 5 %%%%%%%%%%%%%%%%%%%%%%%%%%%%%%%%%%%%%%
\begin{figure}%[!ht]
\begin{center}
\includegraphics[width=8 cm ]{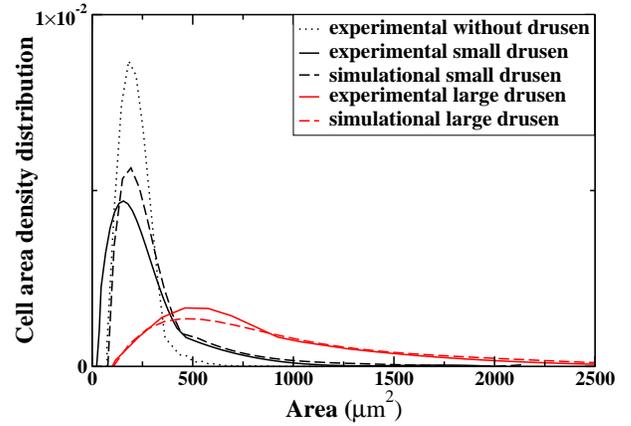} %&
\end{center}
\caption{Comparison of cell area density distribution for cells located on and near soft drusen: simulation results (dashed lines) vs. experimental data (solid lines), and small drusen (black) vs. large drusen (red).  Cells within 100 $\mu m$ from the centers of the drusen were measured.  Ten replica simulations using $J=4$, $\kappa=0.2$ and $T=2$ were performed to obtain each of the area distributions (mean distribution is shown) associated with a small druse of radius $52\;\mu$m (dashed black line),  and a large druse of radius $87\;\mu$m (dashed red line).  Eight small drusen of radius $r_D$ between $40$ and  $55$ $\mu m$ (solid black line), and six larger drusen, $70 < r_D < 90 \;\mu m$ (solid red line), from the same macular region (87 year old male with geographic atrophy, Fig \ref{fig2}C) were measured to obtain the cell area distribution near drusen. Area distribution of a normal macula RPE from an 88 year old female, Fig \ref{fig2}A, is also included for comparison (dotted line).}
\label{comparison}
\end{figure}

Before entering into a detailed analysis of simulation results, we compare the normal RPE to those having soft drusen. We showed the differences in cell areas located on soft drusen and their surroundings, defined as within 100 $\mu m$ from the centers of the drusen, where the tissue distortion can be seen by naked eye (semientras quee Fig. \ref{snap}(g)).  
These results are useful to understand the degree of distortion that can be achieved as a guide to simulations.  Fig. \ref{comparison} shows cell area density distributions for a normal human RPE (dotted line) and the distorted area distribution by soft drusen (solid lines), as well as the distortions by small drusen (black) vs. by large drusen (red).  We found that the mean and variance of the cell area distributions from simulations differed by less than $20 \%$ from experimental values. For small druse,  experimental cell area was $284 \pm 275 \;\mu m^2$; for large druse, experimental cell area was $1010 \pm 1197 \;\mu m^2$.  Comparing to the normal RPE cells ($246 \pm 96 \;\mu m^2$), we see that the small drusen caused cell area distortions significantly less than the large drusen, in this last case not only broaden the area distributions but also increase the cell areas.  \par 

Recall that we simulated soft druse formation by gradually inducing cell apoptosis at the site of soft druse.  The death of about $150$ cells is necessary to simulate on a normal tissue the deterioration induced by a soft druse of $87$ $\mu m$ radius, whereas about $55$ cells die on average to obtain the deterioration induced by soft drusen about $40$ $\mu m$ radius.  For these druse sizes, our model established that triple the number of cells are killed to generate the deterioration of a druse that is only double in size.  This mismatch predicts that the dynamics of soft druse formation play an important role in generating the morphological distortion. In the next section extent and irregularity of damaged epithelia by drusen of different sizes were studied analyzing regularity of Voronoi domain areas derived from distorted RPE cells.

\subsection{Voronoi domain analysis}

\begin{figure}%[!ht]
\begin{center}
\includegraphics[width=8 cm ]{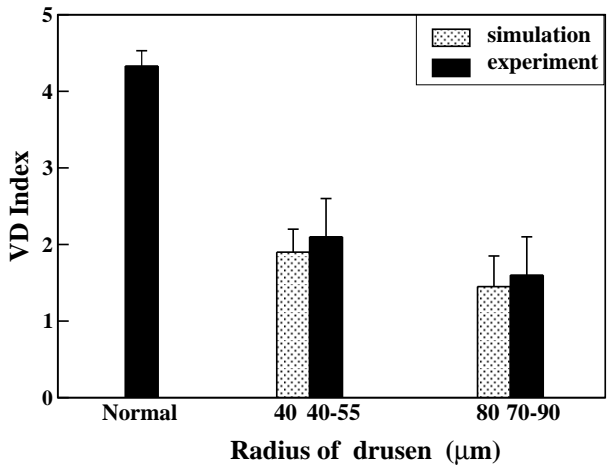} 
\end{center}
\caption{Voronoi domain regularity measure of RPE patterns as a function of druse size, for simulated (white) and experimental (dark) RPE pattern. Normal RPE is as in Fig \ref{fig2}A. Ten simulations were performed for small and big drusen of fixed radius of 40 and 80 $\mu$, respectively. The RPE measured are for tissues within 100  $\mu m$ from the center of soft drusen.}
\label{voronoi}
\end{figure}

We further used a Voronoi-tessellation based spatial domain analysis to describe the uniformity of cell distribution \cite{raven02}. Drusen underneath RPE cells generate tension and  deformation of the tissue extending far from the location of druse. We quantified this damage by measuring the Voronoi regularity indices on and near soft drusen. For this propose, we calculated the centers of mass of the RPE cells of the affected region around each druse. Next, we constructed Voronoi domains (VD) as polygonal shapes that surround the centers, such that any point in a polygon is closer to its generator center than any other center. Finally, we calculated the Voronoi domain regularity index for a tissue sample by dividing the mean Voronoi domain area by the standard deviation of that sample.  Fig. \ref{voronoi} shows the Voronoi domain regularity indices for the normal RPE and different RPE tissues with soft drusen. The indexes were calculated for a circular region 100 $\mu m$ from the center of drusen (Fig. \ref{snap}g). The irregularities of the tissue due to druse increased considerably with the druse size. VD regularity index is above to 4 for a normal RPE and it is  between 1 and 2 for larger drusen. 
If we use the mean and variance of the cell area for normal RPE calculated in a area of radius of 100 $\mu m$ (see section \ref{soft-drusen}), we would obtain $<A>/\Delta A = 2.56$. Therefore, the index of Voronoi allows enlarging of the scale of measurement of the regularity of the tissues, hence is a better measurement for cell packing regularity. 
\par

%%%%%%%%%%%%%%%%%%%%%%%%%%%%%%%%%% FIGURE 6 %%%%%%%%%%%%%%%%%%%%%%%%%%%%%%%%%%%%%%

%%%%%%%%%%%%%%%%%%%%%%%%%%%%%%%%%%%%%%%%%%%%%%%%%%%%%%%%%%%%%%%%%%%%%%%%%%%%%%%%%%%

\subsection{Local anisotropy around drusen}

\begin{figure}%[!ht]
\begin{center}
\includegraphics[width=8 cm ]{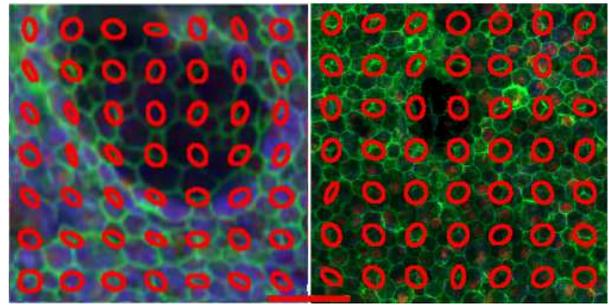} 
\end{center}
\caption{Ellipses generated from tensor texture superimposed on a tissue with a soft druse of $90$ $\mu m$ (left) and $35$ $\mu m$ (right) of radius. Long and short axis ellipses are proportional in length to the largest and smallest eigenvalues and point along the corresponding eigenvector. The subareas are $A(\overline{R})=50\times 50\:\mu m^2$, in which an ellipse is calculated. We particularly see deformation around the druse on the left. The anisotropies measured for these tissues, within 100 $\mu m$ from the center of drusen are $1.8$ and $1.2$, respectively. Practically, the tissues on and around the small drusen keep isotropic whereas the eigenvalues of the texture tensor reveal accumulated tensions on the tissue for big drusen. 
Scale bar = 100 $\mu m$.}
\label{texture}
\end{figure}

Besides cell area distributions and Voronoi-tessellation, we used a texture tensor that measures spatial distributions of pattern
irregularities around drusen. The texture tensor was proposed in \cite{Yi03,Yi03-2} to quantify stored deformation that, in contrast 
to the normal definition of the elastic strain tensor, does not require the details of a reference state. I.E., the texture tensor can be measured on the image directly, without knowledge of the original state of the tissue prior to deformation caused by the druse.  To characterize the texture of a tissue in a given area, we divide this area in subareas $A(\overline{R})$ around the position $\overline{R}$. In each subarea, we list all pairs $(\overline{r}, \overline{r}')$ of positions of neighbor cell centers. From the vector
$\overline{l}= \overline{r} - \overline{r}'$, we construct the tensor   
$\overline{l} \otimes \overline{l}= (l_i l_j )$, where $\otimes$ denotes the standard tensor product and $i$ and $j$ are Cartesian coordinate indices. This tensor, averaged over the number of neighbors of all cells in $A(\overline{R})$, defines the local texture tensor, $M(\overline{R})=\left< l_il_j\right>_{A(\overline{R})}$. $M(\overline{R})$ is symmetric and therefore has two strictly positive eigenvalues. At each  $A(\overline{R})$, we represent $M(\overline{R})$, drawing  an ellipse whose axes are orientated in the direction of its eigenvectors and the dimensions of the axis are proportional to its associated eigenvalue.
Thus, at each  $A(\overline{R})$, the texture tensor field describes an ellipse, which is a measure of the local deformation of cells (the largest axis being in the direction in which cells elongate).  For instance, Fig. \ref{texture} shows an example of $M(\overline{R})$ measured in a RPE with a soft druse. The ellipses result more stretching near the druse and far from them, where here is no local tension, there are circles.\par

The eigenvalues of the texture tensor can be used to characterize texture anisotropy. We defined anisotropy as the ratio of the major axis to the minor axis of the ellipse and averaged over a given area. Note that in this definition the anisotropy is always larger than unity. In fact, around drusen greater than 40 $\mu m$ of radius, the anisotropy was equal to $2.3\pm 0.6$. This value was measured for a circular region of 100 $\mu m$ from the center of soft drusen (the circular region is divided in subareas $A(\overline{R})=50\times 50\;\mu m^2$, in which an ellipse is calculated as we explained previously). Then, the anisotropy was reduced to $1.2\pm 0.2$ far from the druse or measured for normal tissues. 
Drusen less than 40 $\mu m$ of radius generated low deformations, and the anisotropy in these cases was $1.3\pm 0.2$. Numerical simulations of soft drusen greater than 40 $\mu m$ of radius generated an anisotropy of $1.7\pm 0.2$. This value is less than the value obtained experimentally due the three dimensional structures of drusen contributing to the anisotropy value and these deformations were not considered in our model.\par 

%%%%%%%%%%%%%%%%%%%%%%%%%%%%%%%%%% FIGURE 7 %%%%%%%%%%%%%%%%%%%%%%%%%%%%%%%%%%%%%%

%%%%%%%%%%%%%%%%%%%%%%%%%%%%%%%%%%%%%%%%%%%%%%%%%%%%%%%%%%%%%%%%%%%%%%%%%%%%%%%%%%%

\subsection{Characterization of epithelial tissue damaged by different processes of localized cell death}
\label{characterization}

\begin{figure}%[!ht]\includegraphics[width=.8\textwidth]
\begin{center}
\includegraphics[width=8 cm ]{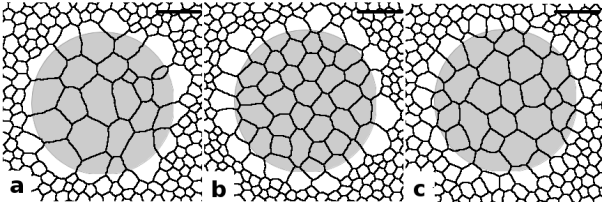} 
\end{center}
\caption{
RPE morphology in the presence of a soft druse of $82\;\mu m$ of radius with cell death caused by:
(a) cell fusions, (b) cell expansions and (c) apoptosis. These snapshots correspond to final RPE images simulated 
following the criteria explained in section \ref{characterization} (shaded regions represent to a druse).  
 Scale bar = $50\;\mu m$.}
\label{processes}
\end{figure}

There are at least three possible mechanisms that may occur during the changes generated on the RPE due to drusen: apoptosis followed by purse-string epithelial repair, cell fusions which lead to bi- tri- up to five nucleate cells and
cell expansions that can reach cell areas five times larger that their original areas followed by the death of neighbor cells due to an excessive compression. The former was considered as the mechanism of cell death and repair of tissue in our previous simulations. In this section, we presented computer simulations validating this process as the main mechanism through the characterization of epithelial tissue modified by the different process mentioned. 
We compared our previous simulations of apoptosis with simulations where the process of localized cell death was modified in two distinct ways: (a) Cell fusion modeled by setting the same ID value for the chosen neighbor cells at random on a lattice localized area (see section \ref{model}); (b) Cell expansion was allowed to reach five times its area while its first, second and third neighboring cells shrank from original area to 20\%, 15\%  and 5\%, respectively. As consequence of the compression, a neighbor cell constricted more than 20\% of its original area, it died enforcing $A_{dead\;cell\;0}=0$ (see Eq. (\ref{energy})).  Both (a) and (b) patterns evolve according to Eqs. (\ref{energy}) and (\ref{prob}) until the systems reach their lowest-energy configurations. We chosen a soft druse location and certain size in each pattern and a cell was chosen at random from the soft druse area and assigned undergo: (a) fusion with a neighbor cell chosen also at random or (b) expansion. The simulations started with normal tissues and these processes were repeated in each tissue separately from within the soft druse area, until the patterns tried to reach similar cell morphology to that observed experimentally. The fusion was limited to a maximum of ten cells, that it is the maximum number of nuclei observed in RPE cells. Fig. \ref{processes} shows the different patterns reached by localized cell death due to: (a) cell fusions, (b) expansions and (c) apoptosis. The final patterns are very different from each other. There are a few cells that they didn't undergo fusion keeping their original size in (a), whereas that the cell expansions generated so uniformity on the (b) pattern, the last one (c) seems more realistic (see left panel in Fig. \ref{texture} for a comparison). 
 
%The (a) and (b) patterns are disctint from the (c) and the experimental pattern (d). (These last tissues are similar as it previously discussed in this work). \par 
%and (d) a real soft druse.
%%%%%%%%%%%%%%%%%%%%%%%%%%%%%%%%%% FIGURE 8 %%%%%%%%%%%%%%%%%%%%%%%%%%%%%%%%%%%%%%

%%%%%%%%%%%%%%%%%%%%%%%%%%%%%%%%%%%%%%%%%%%%%%%%%%%%%%%%%%%%%%%%%%%%%%%%%%%%%%%%%%%

\begin{figure}%[!ht]
\begin{center}
\includegraphics[width=8 cm ]{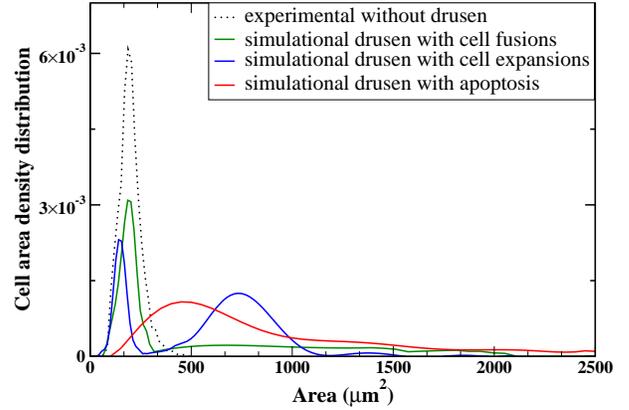} 
\end{center}
\caption{Comparison of cell area density distribution for cells located on and near soft drusen of $82\;\mu m$ of radius with cell death caused
by: cell fusions (green line), expansions (blue line) and apoptosis (red line). This last distribution is similar to the experimental data obtained for big drusen (see Fig. \ref{comparison}) revealing the apoptosis process followed by a purse-string as the main mechanism of localized cell death. Cells within 100 $\mu m$ from the centers of the drusen were measured.  Ten replica simulations using $J=4$, $\kappa=0.2$ and $T=2$ were performed to obtain each of the area distributions (mean distribution is shown) from a normal macula RPE which area distribution is also is also included for comparison (dotted line).}
\label{comparison2}
\end{figure}

For a better comparison between these three processes, in the Fig. \ref{comparison2} cell area density distributions located on and near soft drusen are shown. We found that the distributions due to cell fusion (green line) and expansion (blue line) mechanisms are completely different to the distribution produced by apoptosis and elongation of neighbor cells (red line). This last distribution is similar to experimental data showed in Fig. \ref{comparison}. The constrain of a maximum of cell area in the expansion process generates a distribution with two peaks: the first one is due to the cell compressions and it is on the left of the peak of the distribution of the normal RPE used to start the simulations (dotted line) and the other one is due to the constrain cell maximum area. In contrast, the constrain for the cell fusions is softer (limited to a maximum of ten cells) and it did not produce an area distribution with two peaks. However, this distribution did not fit to the experimental results. We also calculated the
Voronoi indices and the anisotropies through the textor tensor's eigenvalues and obtained from
simulated drusen of $82\;\mu m$ of radius: $1.8 \pm 0.2$ and $1.7 \pm 0.2$ for cell fusions, $2.0 \pm 0.2$ and
$1.2 \pm 0.2$ for cell expansions and $1.6 \pm 0.2$ and $1.7 \pm 0.2$ for apoptosis, respectively. Simulated drusen
from cell expansions followed by compression and cell death did not generate large
irregularities in the tissue whereas similar anisotropy values were found for drusen obtained
from cell fusions and apoptosis but their cell area density distributions were very different located on and near
drusen as we showed in Fig. \ref{comparison2}.
These three processes could be combined, but given the unknowledged on their intervention grades the results shown in this section allowed us to consider the apoptosis followed by a purse-string as the main mechanism of localized cell death.\par

%Starting from a normal
%PRE pattern
%for the tissue modified in the three distinct ways previously :

%a normal human RPE (dotted line) and the
%distorted area distribution by soft drusen (solid lines), as well as the distortions by small
%drusen (black) vs. by large drusen (red). We found that
%simulation results (dashed lines) vs. experimental data (solid lines), and small drusen (black) vs.
%large drusen (red). Cells within 100 μm from the centers of the drusen were measured.
%\par 

%(d) Image of a soft druse from the macular region of a eye with AMD.
%computer simulations correponding to 
%%%%%%%%%%%%%%%%%%%%%%%%%%%%%%%%%% FIGURE 9 %%%%%%%%%%%%%%%%%%%%%%%%%%%%%%%%%%%%%%

%%%%%%%%%%%%%%%%%%%%%%%%%%%%%%%%%%%%%%%%%%%%%%%%%%%%%%%%%%%%%%%%%%%%%%%%%%%%%%%%%Converting genetic network oscillations 

\section{Discussion and Conclusions}

Our aim was to understand the morphology of RPE layer when damaged by different sizes of drusen. Our motivations derive from promising advances in imaging human RPE cells with {\em in vivo}, real-time, adaptive optics scanning devices \cite{Duncan, Roorda}.\par

We asked how cell pattern and tiling arise in living organisms, using knowledge of fundamental properties of cell adhesion.  Despite cell complexity, cell shape is controlled by surface tension to a large extent.  The control of cell shape involves an organization of the plasma membrane into an effective surface of contact with extracellular environment, including contacting cells, extracellular matrix, and medium. In the context of epithelium, the contact surfaces are between neighboring cells bound by various adhesion complexes.  We developed an RPE model based on the Cellular Potts model, which has characteristics sufficient to study the morphology of RPE and alterations caused by drusen. This model includes the essential properties of epithelium, i.e., cell-cell adhesion, cortical tension, and cell area elasticity. This last property is due to the constant volume of an epithelial cell, limiting the size each cell can stretch into in two dimensions.  Although the plasma membrane constitutes a significant reservoir of molecules that are readily available for change in its surface of contact with other cells, cell volumes are constrained. We have reproduced packing geometries observed in human RPE corresponding to our model of local minima in an energy function which forces are balanced.\par

The RPE cells generally do not divide in the adult. However, there is a gradual loss of cells with increasing age and an increase in size of those remaining to compensate for this loss in order to maintain the integrity of the barrier. RPE cell death can cause photoreceptor  degeneration and the development geographic atrophy \cite{Holz2004}. To examine how normal cell packing geometries respond to injured tissue, in our simulations we induced cell apoptosis to create a defect in the tissue integrity, and then allowed the adjacent cells to expand to fill the space previously occupied by the dead cells. Thereby, we reproduced local deformations due to defects in normal human RPE using our computational RPE model. The adjacent cells elongated toward the gap, forming a focus for small tissue injuries and irregular patterns for large injuries. Soft drusen are often found under these injuries in real cadaver eyes that had been diagnosed with AMD.\par

The irregularity of tissues produced by soft drusen was measured using cell area distributions, Voronoi domain analysis and texture tensor. The area distributions for cells on and near the soft drusen reveal a clear increase of their sizes compared to normal cells, due to the localized cell death promoted by the drusen. Our model predicts a localized death of about 150 cells induced by a druse of $87\;\mu m$ of radius and 55 cells for a druse of $40\;\mu m$ of radius. These numbers far exceed the cell numbers covering the drusen (less than thirty cells were observed on small and big drusen). To describe the uniformity of cell distribution and the anisotropy of the tissue generated by drusen, the Voronoi domain regularity indices and the texture tensor for matrices deformed by experiment and simulated drusen of different sizes were calculated. The farther away from the druse, the more regular the Voronoi indices and the less anisotropy value revealing a decrease of the tensions.\par
Finally, a characterization of epithelial tissue damaged by different processes of localized cell death was presented in section \ref{characterization}. Different mechanisms may occur during the changes generated on the RPE due to drusen: cell death followed by a purse-string mechanism, cell fusions and cell expansions. We showed that the first process is the predominant mechanism to characterize tissues damaged by drusen.\par
In summary, simulations of our RPE morphogenesis model quantitatively reproduced deformations of RPE morphology in the presence of drusen, suggesting that a purse-string mechanism is sufficient to explain how RPE heals cell loss caused by drusen-damage.  Our results also suggest that soft drusen induce cell stretching like preponderant mechanism. Taken together, our integrated experimental and modeling study has allowed us to study morphological changes the RPE sheet associated with drusen, and suggest a plausible mechanism of how RPE repair its cell loss.  This model 
 has provided a physical description of the epithelium remodeling and morphological change.  This is exciting because it suggests the possibility that we could use RPE morphology analysis as a diagnostic and prognostic tool.  Non-invasive imaging of the living human eye may soon provide comparable morphometric data on the human RPE. We speculate further possible significance: knowledge of a particular morpho-type may be used to predict a response to a given drug and then to assess the response.

%\section{Bibliography}
%\bibliographystyle{unsrt}  
\bibliography{bibpercola(1)}

\begin{thebibliography}{63}
\expandafter\ifx\csname natexlab\endcsname\relax\def\natexlab#1{#1}\fi
\expandafter\ifx\csname bibnamefont\endcsname\relax
  \def\bibnamefont#1{#1}\fi
\expandafter\ifx\csname bibfnamefont\endcsname\relax
  \def\bibfnamefont#1{#1}\fi
\expandafter\ifx\csname citenamefont\endcsname\relax
  \def\citenamefont#1{#1}\fi
\expandafter\ifx\csname url\endcsname\relax
  \def\url#1{\texttt{#1}}\fi
\expandafter\ifx\csname urlprefix\endcsname\relax\def\urlprefix{URL }\fi
\providecommand{\bibinfo}[2]{#2}
\providecommand{\eprint}[2][]{\url{#2}}

\bibitem[{\citenamefont{Ardeljan and Chan}(in press, 2013)}]{ardeljan2013}
\bibinfo{author}{\bibfnamefont{D.}~\bibnamefont{Ardeljan}} \bibnamefont{and}
  \bibinfo{author}{\bibfnamefont{C.}~\bibnamefont{Chan}},
  \bibinfo{journal}{Prog. Ret. Eye Res.} pp. \bibinfo{pages}{1--22}
  (\bibinfo{year}{in press, 2013}).

\bibitem[{\citenamefont{Coleman et~al.}(2008)\citenamefont{Coleman, Chan,
  Ferris, and Chew}}]{Coleman2008}
\bibinfo{author}{\bibfnamefont{H.~R.} \bibnamefont{Coleman}},
  \bibinfo{author}{\bibfnamefont{C.~C.} \bibnamefont{Chan}},
  \bibinfo{author}{\bibfnamefont{F.~L.} \bibnamefont{Ferris}},
  \bibnamefont{and} \bibinfo{author}{\bibfnamefont{E.~Y.} \bibnamefont{Chew}},
  \bibinfo{journal}{Lancet} \textbf{\bibinfo{volume}{372}},
  \bibinfo{pages}{1835} (\bibinfo{year}{2008}).

\bibitem[{\citenamefont{Bird and et. al.}(1995)}]{cite3Rabbit07}
\bibinfo{author}{\bibfnamefont{A.~C.} \bibnamefont{Bird}} \bibnamefont{and}
  \bibinfo{author}{\bibnamefont{et. al.}}, \bibinfo{journal}{Surv. Ophthalmol.}
  \textbf{\bibinfo{volume}{39}}, \bibinfo{pages}{367} (\bibinfo{year}{1995}).

\bibitem[{\citenamefont{Congdon et~al.}(in press, 2013)\citenamefont{Congdon,
  O’Colmain, Klaver, Klein, Munoz, Friedman, Kempen, Taylor, and
  Mitchell}}]{Congdon2004}
\bibinfo{author}{\bibfnamefont{N.}~\bibnamefont{Congdon}},
  \bibinfo{author}{\bibfnamefont{B.}~\bibnamefont{O’Colmain}},
  \bibinfo{author}{\bibfnamefont{C.~C.} \bibnamefont{Klaver}},
  \bibinfo{author}{\bibfnamefont{R.}~\bibnamefont{Klein}},
  \bibinfo{author}{\bibfnamefont{B.}~\bibnamefont{Munoz}},
  \bibinfo{author}{\bibfnamefont{D.~S.} \bibnamefont{Friedman}},
  \bibinfo{author}{\bibfnamefont{J.}~\bibnamefont{Kempen}},
  \bibinfo{author}{\bibfnamefont{H.~R.} \bibnamefont{Taylor}},
  \bibnamefont{and} \bibinfo{author}{\bibfnamefont{P.}~\bibnamefont{Mitchell}},
  \bibinfo{journal}{Arch. Ophthalmol.} \textbf{\bibinfo{volume}{122}},
  \bibinfo{pages}{477} (\bibinfo{year}{in press, 2013}).

\bibitem[{\citenamefont{Pascolini et~al.}(2004)\citenamefont{Pascolini,
  Mariotti, Pokharel, Pararajasegaram, Etya'ale, N\'egrel, and
  Resnikoff}}]{Pascolini2004}
\bibinfo{author}{\bibfnamefont{D.}~\bibnamefont{Pascolini}},
  \bibinfo{author}{\bibfnamefont{S.~P.} \bibnamefont{Mariotti}},
  \bibinfo{author}{\bibfnamefont{G.~P.} \bibnamefont{Pokharel}},
  \bibinfo{author}{\bibfnamefont{R.}~\bibnamefont{Pararajasegaram}},
  \bibinfo{author}{\bibfnamefont{D.}~\bibnamefont{Etya'ale}},
  \bibinfo{author}{\bibfnamefont{A.~D.} \bibnamefont{N\'egrel}},
  \bibnamefont{and}
  \bibinfo{author}{\bibfnamefont{S.}~\bibnamefont{Resnikoff}},
  \bibinfo{journal}{Ophthalmic Epidemiol.} \textbf{\bibinfo{volume}{11}},
  \bibinfo{pages}{67} (\bibinfo{year}{2004}).

\bibitem[{\citenamefont{Ngai et~al.}(2011)\citenamefont{Ngai, Stocks, Sparrow,
  Patel, Rumley, Lowe, Davey~Smith, and Ben-Shlomo}}]{citetoMiller20}
\bibinfo{author}{\bibfnamefont{L.~Y.} \bibnamefont{Ngai}},
  \bibinfo{author}{\bibfnamefont{N.}~\bibnamefont{Stocks}},
  \bibinfo{author}{\bibfnamefont{J.~M.} \bibnamefont{Sparrow}},
  \bibinfo{author}{\bibfnamefont{R.}~\bibnamefont{Patel}},
  \bibinfo{author}{\bibfnamefont{A.}~\bibnamefont{Rumley}},
  \bibinfo{author}{\bibfnamefont{G.}~\bibnamefont{Lowe}},
  \bibinfo{author}{\bibfnamefont{G.}~\bibnamefont{Davey~Smith}},
  \bibnamefont{and}
  \bibinfo{author}{\bibfnamefont{Y.}~\bibnamefont{Ben-Shlomo}},
  \bibinfo{journal}{Eye} \textbf{\bibinfo{volume}{25}}, \bibinfo{pages}{784}
  (\bibinfo{year}{2011}).

\bibitem[{\citenamefont{van Leeuwen et~al.}(2003)\citenamefont{van Leeuwen,
  Klaver, Vingerling, Hoffman, and de~Jong}}]{cite1Miller10}
\bibinfo{author}{\bibfnamefont{R.}~\bibnamefont{van Leeuwen}},
  \bibinfo{author}{\bibfnamefont{C.~C.~W.} \bibnamefont{Klaver}},
  \bibinfo{author}{\bibfnamefont{J.~R.} \bibnamefont{Vingerling}},
  \bibinfo{author}{\bibfnamefont{A.}~\bibnamefont{Hoffman}}, \bibnamefont{and}
  \bibinfo{author}{\bibfnamefont{P.~T. V.~M.} \bibnamefont{de~Jong}},
  \bibinfo{journal}{Eur. J. Epidemiol.} \textbf{\bibinfo{volume}{18}},
  \bibinfo{pages}{845} (\bibinfo{year}{2003}).

\bibitem[{\citenamefont{Pascolini and Mariotti}(2011)}]{Pascolini2011}
\bibinfo{author}{\bibfnamefont{D.}~\bibnamefont{Pascolini}} \bibnamefont{and}
  \bibinfo{author}{\bibfnamefont{S.~P.} \bibnamefont{Mariotti}},
  \bibinfo{journal}{Br. J. Ophthalmol.}  (\bibinfo{year}{2011}).

\bibitem[{\citenamefont{Friedman et~al.}(2004)\citenamefont{Friedman,
  O’Colmain, Munoz, Tomany, McCarty, de~Jong, Nemesure, Mitchell, and
  Kempen}}]{Friedman2004}
\bibinfo{author}{\bibfnamefont{D.~S.} \bibnamefont{Friedman}},
  \bibinfo{author}{\bibfnamefont{B.~J.} \bibnamefont{O’Colmain}},
  \bibinfo{author}{\bibfnamefont{B.}~\bibnamefont{Munoz}},
  \bibinfo{author}{\bibfnamefont{S.~C.} \bibnamefont{Tomany}},
  \bibinfo{author}{\bibfnamefont{C.}~\bibnamefont{McCarty}},
  \bibinfo{author}{\bibfnamefont{P.~T.} \bibnamefont{de~Jong}},
  \bibinfo{author}{\bibfnamefont{B.}~\bibnamefont{Nemesure}},
  \bibinfo{author}{\bibfnamefont{P.}~\bibnamefont{Mitchell}}, \bibnamefont{and}
  \bibinfo{author}{\bibfnamefont{J.}~\bibnamefont{Kempen}},
  \bibinfo{journal}{Arch. Ophthalmol.} \textbf{\bibinfo{volume}{122}},
  \bibinfo{pages}{564} (\bibinfo{year}{2004}).

\bibitem[{\citenamefont{Brown et~al.}(2009{\natexlab{a}})\citenamefont{Brown,
  Michels, Kaiser, Heier, Sy, and Ianchulev}}]{Brown09}
\bibinfo{author}{\bibfnamefont{D.~M.} \bibnamefont{Brown}},
  \bibinfo{author}{\bibfnamefont{M.}~\bibnamefont{Michels}},
  \bibinfo{author}{\bibfnamefont{P.~K.} \bibnamefont{Kaiser}},
  \bibinfo{author}{\bibfnamefont{J.~S.} \bibnamefont{Heier}},
  \bibinfo{author}{\bibfnamefont{J.~P.} \bibnamefont{Sy}}, \bibnamefont{and}
  \bibinfo{author}{\bibfnamefont{T.}~\bibnamefont{Ianchulev}},
  \bibinfo{journal}{Ophthalmol.} \textbf{\bibinfo{volume}{116}},
  \bibinfo{pages}{57} (\bibinfo{year}{2009}{\natexlab{a}}).

\bibitem[{\citenamefont{Forte et~al.}(2010)\citenamefont{Forte, Cennamo,
  Finelli, Cesarano, D'Amico, de~Crecchio, and Cennamo}}]{Forte10}
\bibinfo{author}{\bibfnamefont{R.}~\bibnamefont{Forte}},
  \bibinfo{author}{\bibfnamefont{G.}~\bibnamefont{Cennamo}},
  \bibinfo{author}{\bibfnamefont{M.}~\bibnamefont{Finelli}},
  \bibinfo{author}{\bibfnamefont{I.}~\bibnamefont{Cesarano}},
  \bibinfo{author}{\bibfnamefont{G.}~\bibnamefont{D'Amico}},
  \bibinfo{author}{\bibfnamefont{G.}~\bibnamefont{de~Crecchio}},
  \bibnamefont{and} \bibinfo{author}{\bibfnamefont{G.}~\bibnamefont{Cennamo}},
  \bibinfo{journal}{Acta Ophthalmol. Scandinavica Found.}
  \textbf{\bibinfo{volume}{88}}, \bibinfo{pages}{e305} (\bibinfo{year}{2010}).

\bibitem[{\citenamefont{Brown et~al.}(2009{\natexlab{b}})\citenamefont{Brown,
  Michels, Kaiser, Heier, Sy, and Ianchulev}}]{Brown2008}
\bibinfo{author}{\bibfnamefont{D.~M.} \bibnamefont{Brown}},
  \bibinfo{author}{\bibfnamefont{M.}~\bibnamefont{Michels}},
  \bibinfo{author}{\bibfnamefont{P.~K.} \bibnamefont{Kaiser}},
  \bibinfo{author}{\bibfnamefont{J.~S.} \bibnamefont{Heier}},
  \bibinfo{author}{\bibfnamefont{J.~P.} \bibnamefont{Sy}}, \bibnamefont{and}
  \bibinfo{author}{\bibfnamefont{T.}~\bibnamefont{Ianchulev}},
  \bibinfo{journal}{Ophthalmol.} \textbf{\bibinfo{volume}{116}},
  \bibinfo{pages}{57} (\bibinfo{year}{2009}{\natexlab{b}}).

\bibitem[{\citenamefont{Martin et~al.}(2011)\citenamefont{Martin, Maguire,
  Ying, Grunwald, Fine, and Jaffe}}]{Martin2011}
\bibinfo{author}{\bibfnamefont{D.~F.} \bibnamefont{Martin}},
  \bibinfo{author}{\bibfnamefont{M.~G.} \bibnamefont{Maguire}},
  \bibinfo{author}{\bibfnamefont{G.~S.} \bibnamefont{Ying}},
  \bibinfo{author}{\bibfnamefont{J.~E.} \bibnamefont{Grunwald}},
  \bibinfo{author}{\bibfnamefont{S.~L.} \bibnamefont{Fine}}, \bibnamefont{and}
  \bibinfo{author}{\bibfnamefont{G.~J.} \bibnamefont{Jaffe}},
  \bibinfo{journal}{N. Engl. J. Med.} \textbf{\bibinfo{volume}{364}},
  \bibinfo{pages}{1897} (\bibinfo{year}{2011}).

\bibitem[{\citenamefont{Strauss}(2005)}]{williams09}
\bibinfo{author}{\bibfnamefont{O.}~\bibnamefont{Strauss}},
  \bibinfo{journal}{Physiol Rev.} \textbf{\bibinfo{volume}{85}},
  \bibinfo{pages}{845} (\bibinfo{year}{2005}).

\bibitem[{\citenamefont{Song and Dunaief}(2013)}]{Song2013}
\bibinfo{author}{\bibfnamefont{D.}~\bibnamefont{Song}} \bibnamefont{and}
  \bibinfo{author}{\bibfnamefont{J.~L.} \bibnamefont{Dunaief}},
  \bibinfo{journal}{Front. Aging Neurosc.} \textbf{\bibinfo{volume}{5}},
  \bibinfo{pages}{1} (\bibinfo{year}{2013}).

\bibitem[{\citenamefont{Jiang et~al.}(2013, in press)\citenamefont{Jiang, Qi,
  Chrenek, Gardner, Boatright, Grossniklaus, and Nickerson}}]{jiang2013}
\bibinfo{author}{\bibfnamefont{Y.}~\bibnamefont{Jiang}},
  \bibinfo{author}{\bibfnamefont{X.}~\bibnamefont{Qi}},
  \bibinfo{author}{\bibfnamefont{M.~A.} \bibnamefont{Chrenek}},
  \bibinfo{author}{\bibfnamefont{C.}~\bibnamefont{Gardner}},
  \bibinfo{author}{\bibfnamefont{J.~H.} \bibnamefont{Boatright}},
  \bibinfo{author}{\bibfnamefont{H.~E.} \bibnamefont{Grossniklaus}},
  \bibnamefont{and} \bibinfo{author}{\bibfnamefont{J.~E.}
  \bibnamefont{Nickerson}}, \bibinfo{journal}{IOVS}
  \textbf{\bibinfo{volume}{13}}, \bibinfo{pages}{doi:10.1167}
  (\bibinfo{year}{2013, in press}).

\bibitem[{\citenamefont{Sparrow and Boulton}(2005)}]{Sparrow2005}
\bibinfo{author}{\bibfnamefont{J.~R.} \bibnamefont{Sparrow}} \bibnamefont{and}
  \bibinfo{author}{\bibfnamefont{M.}~\bibnamefont{Boulton}},
  \bibinfo{journal}{Exp. Eye Res.} \textbf{\bibinfo{volume}{80}},
  \bibinfo{pages}{595–606} (\bibinfo{year}{2005}).

\bibitem[{\citenamefont{Winkler et~al.}(1999)\citenamefont{Winkler, Boulton,
  Gottsch, and Sternberg}}]{Winkler1999}
\bibinfo{author}{\bibfnamefont{B.~S.} \bibnamefont{Winkler}},
  \bibinfo{author}{\bibfnamefont{M.~E.} \bibnamefont{Boulton}},
  \bibinfo{author}{\bibfnamefont{J.~D.} \bibnamefont{Gottsch}},
  \bibnamefont{and}
  \bibinfo{author}{\bibfnamefont{P.}~\bibnamefont{Sternberg}},
  \bibinfo{journal}{Mol. Vis} \textbf{\bibinfo{volume}{5}}, \bibinfo{pages}{32}
  (\bibinfo{year}{1999}).

\bibitem[{Mil(2009)}]{Millodot}
\emph{\bibinfo{title}{Dictionary of Optometry and Visual Science}}
  (\bibinfo{publisher}{(Butterworth-Heinemann)}, \bibinfo{year}{2009}),
  \bibinfo{edition}{7th} ed.

\bibitem[{\citenamefont{Green}(1999)}]{Green1999}
\bibinfo{author}{\bibfnamefont{W.}~\bibnamefont{Green}}, \bibinfo{journal}{Mol.
  Vis.} \textbf{\bibinfo{volume}{5}}, \bibinfo{pages}{27}
  (\bibinfo{year}{1999}).

\bibitem[{\citenamefont{Buch et~al.}(2005)\citenamefont{Buch, Nielsen, Vinding,
  Jensen, Prause, and la~Cour}}]{Buch2005}
\bibinfo{author}{\bibfnamefont{H.}~\bibnamefont{Buch}},
  \bibinfo{author}{\bibfnamefont{N.~V.} \bibnamefont{Nielsen}},
  \bibinfo{author}{\bibfnamefont{T.}~\bibnamefont{Vinding}},
  \bibinfo{author}{\bibfnamefont{G.~B.} \bibnamefont{Jensen}},
  \bibinfo{author}{\bibfnamefont{J.~U.} \bibnamefont{Prause}},
  \bibnamefont{and} \bibinfo{author}{\bibfnamefont{M.}~\bibnamefont{la~Cour}},
  \bibinfo{journal}{Ophthalmol.} \textbf{\bibinfo{volume}{112}},
  \bibinfo{pages}{787} (\bibinfo{year}{2005}).

\bibitem[{\citenamefont{Rudolf et~al.}(2008)\citenamefont{Rudolf, Clark,
  Chimento, Li, Medeiros, and Curcio}}]{Rudolf2008}
\bibinfo{author}{\bibfnamefont{M.}~\bibnamefont{Rudolf}},
  \bibinfo{author}{\bibfnamefont{M.~E.} \bibnamefont{Clark}},
  \bibinfo{author}{\bibfnamefont{M.~F.} \bibnamefont{Chimento}},
  \bibinfo{author}{\bibfnamefont{C.~M.} \bibnamefont{Li}},
  \bibinfo{author}{\bibfnamefont{N.~E.} \bibnamefont{Medeiros}},
  \bibnamefont{and} \bibinfo{author}{\bibfnamefont{C.~A.}
  \bibnamefont{Curcio}}, \bibinfo{journal}{Ophthalmol. Vis. Sci.}
  \textbf{\bibinfo{volume}{49}}, \bibinfo{pages}{1200} (\bibinfo{year}{2008}).

\bibitem[{\citenamefont{O'Brien et~al.}(2002)\citenamefont{O'Brien, Zegers, and
  Mostov}}]{Mostov02}
\bibinfo{author}{\bibfnamefont{L.~E.} \bibnamefont{O'Brien}},
  \bibinfo{author}{\bibfnamefont{M.~M.~P.} \bibnamefont{Zegers}},
  \bibnamefont{and} \bibinfo{author}{\bibfnamefont{K.~E.}
  \bibnamefont{Mostov}}, \bibinfo{journal}{Nat. Rev. Mol. Cell Biol.}
  \textbf{\bibinfo{volume}{3}}, \bibinfo{pages}{531} (\bibinfo{year}{2002}).

\bibitem[{\citenamefont{Sjostrand and Nilsson}(1964)}]{Nilsson}
\bibinfo{author}{\bibfnamefont{F.~S.} \bibnamefont{Sjostrand}}
  \bibnamefont{and} \bibinfo{author}{\bibfnamefont{S.~E.}
  \bibnamefont{Nilsson}}, \emph{\bibinfo{title}{Single Cell-Based Models in
  Biology and Medicine, Mathematics and Biosciences in Interaction}}
  (\bibinfo{publisher}{Charles C Thomas, Springfield}, \bibinfo{year}{1964}).

\bibitem[{\citenamefont{Mawas}(1953)}]{Mawas}
\bibinfo{author}{\bibfnamefont{J.}~\bibnamefont{Mawas}}, \bibinfo{journal}{Ann
  Ocul (Paris)} \textbf{\bibinfo{volume}{186}}, \bibinfo{pages}{488}
  (\bibinfo{year}{1953}).

\bibitem[{\citenamefont{H.}(1969)}]{Moyer}
\bibinfo{author}{\bibfnamefont{M.~F.} \bibnamefont{H.}}, \bibinfo{journal}{UCLA
  Forum Med Sci} \textbf{\bibinfo{volume}{8}}, \bibinfo{pages}{1}
  (\bibinfo{year}{1969}).

\bibitem[{\citenamefont{Farquhar and Palade}(1963)}]{Farquhar}
\bibinfo{author}{\bibfnamefont{M.}~\bibnamefont{Farquhar}} \bibnamefont{and}
  \bibinfo{author}{\bibfnamefont{G.}~\bibnamefont{Palade}}, \bibinfo{journal}{J
  Cell Biol.} \textbf{\bibinfo{volume}{17}}, \bibinfo{pages}{375}
  (\bibinfo{year}{1963}).

\bibitem[{\citenamefont{Bertet et~al.}(2004)\citenamefont{Bertet, Sulak, and
  Lecuit}}]{Julicher2007:3}
\bibinfo{author}{\bibfnamefont{C.}~\bibnamefont{Bertet}},
  \bibinfo{author}{\bibfnamefont{L.}~\bibnamefont{Sulak}}, \bibnamefont{and}
  \bibinfo{author}{\bibfnamefont{T.}~\bibnamefont{Lecuit}},
  \bibinfo{journal}{Nature} \textbf{\bibinfo{volume}{429}},
  \bibinfo{pages}{667} (\bibinfo{year}{2004}).

\bibitem[{\citenamefont{Otani et~al.}(2006)\citenamefont{Otani, Ichii, Aono,
  and Takeichi}}]{Julicher2007:10}
\bibinfo{author}{\bibfnamefont{T.}~\bibnamefont{Otani}},
  \bibinfo{author}{\bibfnamefont{T.}~\bibnamefont{Ichii}},
  \bibinfo{author}{\bibfnamefont{S.}~\bibnamefont{Aono}}, \bibnamefont{and}
  \bibinfo{author}{\bibfnamefont{M.}~\bibnamefont{Takeichi}},
  \bibinfo{journal}{J. Cell Biol.} \textbf{\bibinfo{volume}{175}},
  \bibinfo{pages}{135–146} (\bibinfo{year}{2006}).

\bibitem[{\citenamefont{Franke et~al.}(2005)\citenamefont{Franke, Montague, and
  Kiehart}}]{Julicher2007:11}
\bibinfo{author}{\bibfnamefont{J.~D.} \bibnamefont{Franke}},
  \bibinfo{author}{\bibfnamefont{R.~A.} \bibnamefont{Montague}},
  \bibnamefont{and} \bibinfo{author}{\bibfnamefont{D.~P.}
  \bibnamefont{Kiehart}}, \bibinfo{journal}{Curr. Biol.}
  \textbf{\bibinfo{volume}{15}}, \bibinfo{pages}{2208–2221}
  (\bibinfo{year}{2005}).

\bibitem[{\citenamefont{Starnes et~al.}(2016)\citenamefont{Starnes, Huisinh,
  Mcgwin~Jr., Sloan, Ablonczy, Smith, Curcio, and Ach}}]{Starnes2016}
\bibinfo{author}{\bibfnamefont{A.~C.} \bibnamefont{Starnes}},
  \bibinfo{author}{\bibfnamefont{C.}~\bibnamefont{Huisinh}},
  \bibinfo{author}{\bibfnamefont{G.}~\bibnamefont{Mcgwin~Jr.}},
  \bibinfo{author}{\bibfnamefont{K.~R.} \bibnamefont{Sloan}},
  \bibinfo{author}{\bibfnamefont{Z.}~\bibnamefont{Ablonczy}},
  \bibinfo{author}{\bibfnamefont{T.}~\bibnamefont{Smith}},
  \bibinfo{author}{\bibfnamefont{C.~A.} \bibnamefont{Curcio}},
  \bibnamefont{and} \bibinfo{author}{\bibfnamefont{T.}~\bibnamefont{Ach}},
  \bibinfo{journal}{Visual Neuroscience} \textbf{\bibinfo{volume}{3}},
  \bibinfo{pages}{e001} (\bibinfo{year}{2016}).

\bibitem[{\citenamefont{M. et~al.}(1981)\citenamefont{M., G., P., and
  B.}}]{Odell}
\bibinfo{author}{\bibfnamefont{O.~G.} \bibnamefont{M.}},
  \bibinfo{author}{\bibfnamefont{O.}~\bibnamefont{G.}},
  \bibinfo{author}{\bibfnamefont{A.}~\bibnamefont{P.}}, \bibnamefont{and}
  \bibinfo{author}{\bibfnamefont{B.}~\bibnamefont{B.}}, \bibinfo{journal}{Dev.
  Biol.} \textbf{\bibinfo{volume}{85}}, \bibinfo{pages}{446}
  (\bibinfo{year}{1981}).

\bibitem[{\citenamefont{Sherratt and Murray}(1990)}]{Murray}
\bibinfo{author}{\bibfnamefont{J.~A.} \bibnamefont{Sherratt}} \bibnamefont{and}
  \bibinfo{author}{\bibfnamefont{J.~D.} \bibnamefont{Murray}},
  \bibinfo{journal}{Proc. R. Soc. Lond. B} \textbf{\bibinfo{volume}{241}},
  \bibinfo{pages}{29} (\bibinfo{year}{1990}).

\bibitem[{\citenamefont{McDougall et~al.}(2006)\citenamefont{McDougall, Dallon,
  Sherratt, and Maini}}]{McDougall}
\bibinfo{author}{\bibfnamefont{S.}~\bibnamefont{McDougall}},
  \bibinfo{author}{\bibfnamefont{J.}~\bibnamefont{Dallon}},
  \bibinfo{author}{\bibfnamefont{J.}~\bibnamefont{Sherratt}}, \bibnamefont{and}
  \bibinfo{author}{\bibfnamefont{P.}~\bibnamefont{Maini}},
  \bibinfo{journal}{Phil. Trans. R. Soc. A} \textbf{\bibinfo{volume}{364}},
  \bibinfo{pages}{1385} (\bibinfo{year}{2006}).

\bibitem[{\citenamefont{Posta and Chou}(2010)}]{Posta}
\bibinfo{author}{\bibfnamefont{F.}~\bibnamefont{Posta}} \bibnamefont{and}
  \bibinfo{author}{\bibfnamefont{T.}~\bibnamefont{Chou}}, \bibinfo{journal}{J.
  Theor. Biol.} \textbf{\bibinfo{volume}{266}}, \bibinfo{pages}{70}
  (\bibinfo{year}{2010}).

\bibitem[{\citenamefont{Shin et~al.}(2010)\citenamefont{Shin, Rath, Zebisch,
  Choo, Kolch, and H.}}]{Shin}
\bibinfo{author}{\bibfnamefont{S.~Y.} \bibnamefont{Shin}},
  \bibinfo{author}{\bibfnamefont{O.}~\bibnamefont{Rath}},
  \bibinfo{author}{\bibfnamefont{A.}~\bibnamefont{Zebisch}},
  \bibinfo{author}{\bibfnamefont{S.~M.} \bibnamefont{Choo}},
  \bibinfo{author}{\bibfnamefont{W.}~\bibnamefont{Kolch}}, \bibnamefont{and}
  \bibinfo{author}{\bibfnamefont{C.~K.} \bibnamefont{H.}},
  \bibinfo{journal}{Cancer Res.} \textbf{\bibinfo{volume}{70}},
  \bibinfo{pages}{6715} (\bibinfo{year}{2010}).

\bibitem[{\citenamefont{Davidson}(2002)}]{Merks09:12}
\bibinfo{author}{\bibfnamefont{E.~H.} \bibnamefont{Davidson}},
  \bibinfo{journal}{Science} p. \bibinfo{pages}{1669–1678}
  (\bibinfo{year}{2002}).

\bibitem[{\citenamefont{Glazier et~al.}(2007)\citenamefont{Glazier, Balter, and
  Pop{\l}awski}}]{Merks09:16}
\bibinfo{author}{\bibfnamefont{J.~A.} \bibnamefont{Glazier}},
  \bibinfo{author}{\bibfnamefont{A.}~\bibnamefont{Balter}}, \bibnamefont{and}
  \bibinfo{author}{\bibfnamefont{N.~J.} \bibnamefont{Pop{\l}awski}},
  \emph{\bibinfo{title}{Single Cell-Based Models in Biology and Medicine,
  Mathematics and Bio- sciences in Interaction}}
  (\bibinfo{publisher}{Birkha\"user, Basel, Switzerland},
  \bibinfo{year}{2007}).

\bibitem[{\citenamefont{Merks and Perryn}(2008)}]{Merks09:33}
\bibinfo{author}{\bibfnamefont{R.~M.~H.} \bibnamefont{Merks}} \bibnamefont{and}
  \bibinfo{author}{\bibfnamefont{J.~A.} \bibnamefont{Perryn},
  \bibfnamefont{E.~D.~Glazier}}, \bibinfo{journal}{PLoS Comput. Biol.}
  \textbf{\bibinfo{volume}{4}}, \bibinfo{pages}{e1000163}
  (\bibinfo{year}{2008}).

\bibitem[{\citenamefont{K\"afer et~al.}(2007)\citenamefont{K\"afer, Hayashi,
  Mar\'ee, Carthew, and Graner}}]{Merks09:21}
\bibinfo{author}{\bibfnamefont{J.}~\bibnamefont{K\"afer}},
  \bibinfo{author}{\bibfnamefont{T.}~\bibnamefont{Hayashi}},
  \bibinfo{author}{\bibfnamefont{A.~F.~M.} \bibnamefont{Mar\'ee}},
  \bibinfo{author}{\bibfnamefont{R.~W.} \bibnamefont{Carthew}},
  \bibnamefont{and} \bibinfo{author}{\bibfnamefont{F.}~\bibnamefont{Graner}},
  \bibinfo{journal}{Proc. Natl. Acad. Sci. U.S.A.}
  \textbf{\bibinfo{volume}{104}}, \bibinfo{pages}{18549–54}
  (\bibinfo{year}{2007}).

\bibitem[{\citenamefont{Grieneisen et~al.}(2007)\citenamefont{Grieneisen, Xu,
  Mar\'ee, Hogeweg, and Scheres}}]{Merks09:18}
\bibinfo{author}{\bibfnamefont{V.~A.} \bibnamefont{Grieneisen}},
  \bibinfo{author}{\bibfnamefont{J.}~\bibnamefont{Xu}},
  \bibinfo{author}{\bibfnamefont{A.~F.~M.} \bibnamefont{Mar\'ee}},
  \bibinfo{author}{\bibfnamefont{P.}~\bibnamefont{Hogeweg}}, \bibnamefont{and}
  \bibinfo{author}{\bibfnamefont{B.}~\bibnamefont{Scheres}},
  \bibinfo{journal}{Nature} \textbf{\bibinfo{volume}{449}},
  \bibinfo{pages}{1008–13} (\bibinfo{year}{2007}).

\bibitem[{\citenamefont{Pop{\l}awski et~al.}(2008)\citenamefont{Pop{\l}awski,
  Shirinifard, Swat, and Glazier}}]{Merks09:38}
\bibinfo{author}{\bibfnamefont{N.~J.} \bibnamefont{Pop{\l}awski}},
  \bibinfo{author}{\bibfnamefont{A.}~\bibnamefont{Shirinifard}},
  \bibinfo{author}{\bibfnamefont{M.}~\bibnamefont{Swat}}, \bibnamefont{and}
  \bibinfo{author}{\bibfnamefont{J.~A.} \bibnamefont{Glazier}},
  \bibinfo{journal}{Math. Biosci. Eng.} \textbf{\bibinfo{volume}{5}},
  \bibinfo{pages}{355} (\bibinfo{year}{2008}).

\bibitem[{\citenamefont{Xu et~al.}(2009)\citenamefont{Xu, Chen, Shadden,
  Marsden, Kamocka, Rosen, and Alber}}]{Merks09:58}
\bibinfo{author}{\bibfnamefont{Z.}~\bibnamefont{Xu}},
  \bibinfo{author}{\bibfnamefont{N.}~\bibnamefont{Chen}},
  \bibinfo{author}{\bibfnamefont{S.~C.} \bibnamefont{Shadden}},
  \bibinfo{author}{\bibfnamefont{J.~E.} \bibnamefont{Marsden}},
  \bibinfo{author}{\bibfnamefont{M.~M.} \bibnamefont{Kamocka}},
  \bibinfo{author}{\bibfnamefont{E.~D.} \bibnamefont{Rosen}}, \bibnamefont{and}
  \bibinfo{author}{\bibfnamefont{M.~S.} \bibnamefont{Alber}},
  \bibinfo{journal}{PLoS Comput. Biol.} p. \bibinfo{pages}{769–779}
  (\bibinfo{year}{2009}).

\bibitem[{\citenamefont{Guirao et~al.}(2015)\citenamefont{Guirao, Rigaud,
  Bosveld, Bailles, L\'opez-Gay, Ishihara, Sugimura, Graner, and
  Bella\"ıche}}]{Graner2015}
\bibinfo{author}{\bibfnamefont{B.}~\bibnamefont{Guirao}},
  \bibinfo{author}{\bibfnamefont{S.~U.} \bibnamefont{Rigaud}},
  \bibinfo{author}{\bibfnamefont{F.}~\bibnamefont{Bosveld}},
  \bibinfo{author}{\bibfnamefont{A.}~\bibnamefont{Bailles}},
  \bibinfo{author}{\bibfnamefont{J.}~\bibnamefont{L\'opez-Gay}},
  \bibinfo{author}{\bibfnamefont{S.}~\bibnamefont{Ishihara}},
  \bibinfo{author}{\bibfnamefont{K.}~\bibnamefont{Sugimura}},
  \bibinfo{author}{\bibfnamefont{F.}~\bibnamefont{Graner}}, \bibnamefont{and}
  \bibinfo{author}{\bibfnamefont{Y.}~\bibnamefont{Bella\"ıche}},
  \bibinfo{journal}{eLife} \textbf{\bibinfo{volume}{4}},
  \bibinfo{pages}{e08519} (\bibinfo{year}{2015}).

\bibitem[{\citenamefont{Anderson et~al.}(2007)\citenamefont{Anderson, Chaplain,
  and Rejniak}}]{25ofGraner07}
\bibinfo{editor}{\bibfnamefont{A.~R.~A.} \bibnamefont{Anderson}},
  \bibinfo{editor}{\bibfnamefont{M.~A.~J.} \bibnamefont{Chaplain}},
  \bibnamefont{and} \bibinfo{editor}{\bibfnamefont{K.~A.}
  \bibnamefont{Rejniak}}, eds., \emph{\bibinfo{title}{Single-Cell-Based Models
  in Biology and Medicine}} (\bibinfo{publisher}{(Birkh\"auser-Verlag, Basel)},
  \bibinfo{year}{2007}).

\bibitem[{\citenamefont{Mombach et~al.}(1995)\citenamefont{Mombach, Glazier,
  Raphael, and Zajac}}]{Glazier95}
\bibinfo{author}{\bibfnamefont{J.~C.} \bibnamefont{Mombach}},
  \bibinfo{author}{\bibfnamefont{J.~A.} \bibnamefont{Glazier}},
  \bibinfo{author}{\bibfnamefont{R.~C.} \bibnamefont{Raphael}},
  \bibnamefont{and} \bibinfo{author}{\bibfnamefont{M.}~\bibnamefont{Zajac}},
  \bibinfo{journal}{Phys. Rev. Lett.} \textbf{\bibinfo{volume}{75}},
  \bibinfo{pages}{2244} (\bibinfo{year}{1995}).

\bibitem[{\citenamefont{Jiang et~al.}(1998)\citenamefont{Jiang, Levine, and
  Glazier}}]{Jiang98}
\bibinfo{author}{\bibfnamefont{Y.}~\bibnamefont{Jiang}},
  \bibinfo{author}{\bibfnamefont{H.}~\bibnamefont{Levine}}, \bibnamefont{and}
  \bibinfo{author}{\bibfnamefont{J.~A.} \bibnamefont{Glazier}},
  \bibinfo{journal}{Biophys. J.} \textbf{\bibinfo{volume}{75}},
  \bibinfo{pages}{2615} (\bibinfo{year}{1998}).

\bibitem[{\citenamefont{Jiang et~al.}(2005)\citenamefont{Jiang,
  Pjesivac-Grbovic, Cantrell, and Freyer}}]{Jiang05}
\bibinfo{author}{\bibfnamefont{Y.}~\bibnamefont{Jiang}},
  \bibinfo{author}{\bibfnamefont{J.}~\bibnamefont{Pjesivac-Grbovic}},
  \bibinfo{author}{\bibfnamefont{C.}~\bibnamefont{Cantrell}}, \bibnamefont{and}
  \bibinfo{author}{\bibfnamefont{J.}~\bibnamefont{Freyer}},
  \bibinfo{journal}{Biophys. J.} \textbf{\bibinfo{volume}{89}},
  \bibinfo{pages}{3873} (\bibinfo{year}{2005}).

\bibitem[{\citenamefont{Shirinifard et~al.}(2012)\citenamefont{Shirinifard,
  Glazier, Swat, Gens, Family, Jian, and Grossniklaus}}]{Jiang2012}
\bibinfo{author}{\bibfnamefont{A.}~\bibnamefont{Shirinifard}},
  \bibinfo{author}{\bibfnamefont{J.~A.} \bibnamefont{Glazier}},
  \bibinfo{author}{\bibfnamefont{M.}~\bibnamefont{Swat}},
  \bibinfo{author}{\bibfnamefont{J.~S.} \bibnamefont{Gens}},
  \bibinfo{author}{\bibfnamefont{F.}~\bibnamefont{Family}},
  \bibinfo{author}{\bibfnamefont{Y.}~\bibnamefont{Jian}}, \bibnamefont{and}
  \bibinfo{author}{\bibfnamefont{H.~E.} \bibnamefont{Grossniklaus}},
  \bibinfo{journal}{PLoS Comput. Biol.} \textbf{\bibinfo{volume}{8}},
  \bibinfo{pages}{e1002440} (\bibinfo{year}{2012}).

\bibitem[{\citenamefont{Lecuit and Lenne}(2007)}]{Lecuit}
\bibinfo{author}{\bibfnamefont{T.}~\bibnamefont{Lecuit}} \bibnamefont{and}
  \bibinfo{author}{\bibfnamefont{P.~F.} \bibnamefont{Lenne}},
  \bibinfo{journal}{Nat. Rev. Mol. Cell Biol.} \textbf{\bibinfo{volume}{8}},
  \bibinfo{pages}{633} (\bibinfo{year}{2007}).

\bibitem[{\citenamefont{Glazier and Graner}(1993)}]{Glazier93}
\bibinfo{author}{\bibfnamefont{J.~A.} \bibnamefont{Glazier}} \bibnamefont{and}
  \bibinfo{author}{\bibfnamefont{F.}~\bibnamefont{Graner}},
  \bibinfo{journal}{Phys. Rev. E} \textbf{\bibinfo{volume}{47}},
  \bibinfo{pages}{2128} (\bibinfo{year}{1993}).

\bibitem[{\citenamefont{Thomas R.~Weikl et~al.}(2009)\citenamefont{Thomas
  R.~Weikl, Asfaw, Krobath, R\'o\.{z}ycki, and Lipowsky}}]{Weikl2009}
\bibinfo{author}{\bibfnamefont{T.~R.} \bibnamefont{Thomas R.~Weikl}},
  \bibinfo{author}{\bibfnamefont{M.}~\bibnamefont{Asfaw}},
  \bibinfo{author}{\bibfnamefont{H.}~\bibnamefont{Krobath}},
  \bibinfo{author}{\bibfnamefont{B.}~\bibnamefont{R\'o\.{z}ycki}},
  \bibnamefont{and} \bibinfo{author}{\bibfnamefont{R.}~\bibnamefont{Lipowsky}},
  \bibinfo{journal}{Soft Matter} \textbf{\bibinfo{volume}{5}},
  \bibinfo{pages}{3213–3224} (\bibinfo{year}{2009}).

\bibitem[{\citenamefont{Binder and Heermann}(1997)}]{16deGraner05}
\bibinfo{author}{\bibfnamefont{K.}~\bibnamefont{Binder}} \bibnamefont{and}
  \bibinfo{author}{\bibfnamefont{D.~W.} \bibnamefont{Heermann}},
  \emph{\bibinfo{title}{Monte Carlo Simulation in Statistical Physics}}
  (\bibinfo{publisher}{(Springer, Berlin)}, \bibinfo{year}{1997}),
  \bibinfo{edition}{3rd} ed.

\bibitem[{\citenamefont{Sims et~al.}(1992)\citenamefont{Sims, Karp, and
  Ingber}}]{Ingber:92}
\bibinfo{author}{\bibfnamefont{J.}~\bibnamefont{Sims}},
  \bibinfo{author}{\bibfnamefont{S.}~\bibnamefont{Karp}}, \bibnamefont{and}
  \bibinfo{author}{\bibfnamefont{D.~E.} \bibnamefont{Ingber}},
  \bibinfo{journal}{J. Cell Sci.} \textbf{\bibinfo{volume}{103}},
  \bibinfo{pages}{1215} (\bibinfo{year}{1992}).

\bibitem[{\citenamefont{A. et~al.}(2001)\citenamefont{A., A., and
  P.}}]{Jacinto}
\bibinfo{author}{\bibfnamefont{J.}~\bibnamefont{A.}},
  \bibinfo{author}{\bibfnamefont{M.-A.} \bibnamefont{A.}}, \bibnamefont{and}
  \bibinfo{author}{\bibfnamefont{M.}~\bibnamefont{P.}}, \bibinfo{journal}{Nat.
  Cell Biol.} \textbf{\bibinfo{volume}{3}}, \bibinfo{pages}{E117}
  (\bibinfo{year}{2001}).

\bibitem[{\citenamefont{M. et~al.}(1993)\citenamefont{M., P., and S.}}]{Bement}
\bibinfo{author}{\bibfnamefont{B.~W.} \bibnamefont{M.}},
  \bibinfo{author}{\bibfnamefont{F.}~\bibnamefont{P.}}, \bibnamefont{and}
  \bibinfo{author}{\bibfnamefont{M.~M.} \bibnamefont{S.}}, \bibinfo{journal}{J.
  Cell Biol.} \textbf{\bibinfo{volume}{121}}, \bibinfo{pages}{565}
  (\bibinfo{year}{1993}).

\bibitem[{\citenamefont{D. and A.}(1995)}]{Nobes}
\bibinfo{author}{\bibfnamefont{N.~C.} \bibnamefont{D.}} \bibnamefont{and}
  \bibinfo{author}{\bibfnamefont{H.}~\bibnamefont{A.}},
  \bibinfo{journal}{Cell.} \textbf{\bibinfo{volume}{81}}, \bibinfo{pages}{53}
  (\bibinfo{year}{1995}).

\bibitem[{\citenamefont{Raven and Reese}(2002)}]{raven02}
\bibinfo{author}{\bibfnamefont{M.~A.} \bibnamefont{Raven}} \bibnamefont{and}
  \bibinfo{author}{\bibfnamefont{B.~E.} \bibnamefont{Reese}},
  \bibinfo{journal}{J. Comp. Neurol.} \textbf{\bibinfo{volume}{454}},
  \bibinfo{pages}{168} (\bibinfo{year}{2002}).

\bibitem[{\citenamefont{Aubouy et~al.}(2003)\citenamefont{Aubouy, Jiang,
  Glazier, and Graner}}]{Yi03}
\bibinfo{author}{\bibfnamefont{M.}~\bibnamefont{Aubouy}},
  \bibinfo{author}{\bibfnamefont{Y.}~\bibnamefont{Jiang}},
  \bibinfo{author}{\bibfnamefont{J.~A.} \bibnamefont{Glazier}},
  \bibnamefont{and} \bibinfo{author}{\bibfnamefont{F.}~\bibnamefont{Graner}},
  \bibinfo{journal}{Gran. Matt.} \textbf{\bibinfo{volume}{5}},
  \bibinfo{pages}{67} (\bibinfo{year}{2003}).

\bibitem[{\citenamefont{Asipauskas et~al.}(2003)\citenamefont{Asipauskas,
  Aubouy, Glazier, Graner, and Jiang}}]{Yi03-2}
\bibinfo{author}{\bibfnamefont{M.}~\bibnamefont{Asipauskas}},
  \bibinfo{author}{\bibfnamefont{M.}~\bibnamefont{Aubouy}},
  \bibinfo{author}{\bibfnamefont{J.~A.} \bibnamefont{Glazier}},
  \bibinfo{author}{\bibfnamefont{F.}~\bibnamefont{Graner}}, \bibnamefont{and}
  \bibinfo{author}{\bibfnamefont{Y.}~\bibnamefont{Jiang}},
  \bibinfo{journal}{Gran. Matt.} \textbf{\bibinfo{volume}{5}},
  \bibinfo{pages}{71} (\bibinfo{year}{2003}).

\bibitem[{\citenamefont{Godara et~al.}(2010)\citenamefont{Godara, Dubis,
  Roorda, Duncan, and Carroll}}]{Duncan}
\bibinfo{author}{\bibfnamefont{P.}~\bibnamefont{Godara}},
  \bibinfo{author}{\bibfnamefont{A.~M.} \bibnamefont{Dubis}},
  \bibinfo{author}{\bibfnamefont{A.}~\bibnamefont{Roorda}},
  \bibinfo{author}{\bibfnamefont{J.~L.} \bibnamefont{Duncan}},
  \bibnamefont{and} \bibinfo{author}{\bibfnamefont{J.}~\bibnamefont{Carroll}},
  \bibinfo{journal}{Optom. Vis. Sci.} \textbf{\bibinfo{volume}{87}},
  \bibinfo{pages}{930} (\bibinfo{year}{2010}).

\bibitem[{\citenamefont{Roorda et~al.}(2007)\citenamefont{Roorda, Zhang, and
  Duncan}}]{Roorda}
\bibinfo{author}{\bibfnamefont{A.}~\bibnamefont{Roorda}},
  \bibinfo{author}{\bibfnamefont{Y.}~\bibnamefont{Zhang}}, \bibnamefont{and}
  \bibinfo{author}{\bibfnamefont{J.~L.} \bibnamefont{Duncan}},
  \bibinfo{journal}{Invest. Ophthalmol. Vis. Sci.}
  \textbf{\bibinfo{volume}{48}}, \bibinfo{pages}{2297} (\bibinfo{year}{2007}).

\bibitem[{\citenamefont{Holz et~al.}(2004)\citenamefont{Holz, Pauleikhoff,
  Spaide, and Bird}}]{Holz2004}
\bibinfo{author}{\bibfnamefont{F.~G.} \bibnamefont{Holz}},
  \bibinfo{author}{\bibfnamefont{D.}~\bibnamefont{Pauleikhoff}},
  \bibinfo{author}{\bibfnamefont{R.~F.} \bibnamefont{Spaide}},
  \bibnamefont{and} \bibinfo{author}{\bibfnamefont{A.~C.} \bibnamefont{Bird}},
  \emph{\bibinfo{title}{Age-related macular degeneration}}
  (\bibinfo{publisher}{Springer, Berlin}, \bibinfo{year}{2004}).

\end{thebibliography}

\newpage

\end{document}